\documentclass{article}

\usepackage[preprint]{neurips_2026}
\usepackage{microtype}
\usepackage{graphicx}
\usepackage{subfigure}
\usepackage{booktabs} 
\usepackage{xcolor}
\usepackage{amsmath}
\usepackage{algorithm}
\usepackage{wrapfig}
\usepackage[utf8]{inputenc} 
\usepackage[T1]{fontenc}    
\usepackage{hyperref}       
\usepackage{url}            
\usepackage{booktabs}       
\usepackage{amsfonts}       
\usepackage{nicefrac}       
\usepackage{microtype}      
\usepackage{xcolor}         

\title{Exploring the Alignment of Generation and Understanding in Protein Structure Modeling}

\author{%
  Junde XU$^{1*}$ \quad Yuansheng Huang$^{2,3}$\thanks{Equal contribution.} \quad Zijun Gao$^{1}$ \quad Lihang Liu$^{4}$ \quad Xiaoming Fang$^{4}$ \quad \\ \textbf{Yu Kang} $^{2,3}$ \quad  
  \textbf{Jiezhong Qiu}$^{5}$\thanks{Corresponding author.} \quad \textbf{Pheng-Ann Heng}$^{1}$ \\
    $^{1}$CUHK \quad $^{2}$ College of Pharmaceutical Sciences, ZJU \\ \quad $^{3}$ Sii \quad $^{4}$ Baidu Inc. $^{5}$Hangzhou Institute of Medicine, CAS \\
    \texttt{\url{qiujiezhong@him.cas.cn}}
}

\begin{document}

\maketitle

\begin{abstract}
Understanding and generation are often treated as two separate paradigms in training deep neural networks, despite the fact that both are trained with closely related objectives such as denoising and masked prediction. 
While prior studies have shown that generative models often learn suboptimal representations for understanding tasks in vision, it is less understood whether a similar gap exists in the protein domain. 
In this work, we systematically investigate this question by benchmarking state-of-the-art protein generative models on widely-used protein understanding tasks, and observe that these models exhibit consistently poor performance compared to existing protein encoders.
Furthermore, inspired by the Representation Alignment (REPA) framework~\citep{yu2024representation}, we propose to explicitly align generative protein diffusion models with pretrained protein understanding models during training. 
Experiments on the MotifBench demonstrate that representation alignment significantly improves functional protein generation, boosting the MotifBench score of Protpardelle-1c~\citep{lu2025conditional} from 39.2 to 47.1
, corresponding to a 20\% relative improvement. Our results suggest that representation alignment provides a general and effective mechanism for bridging understanding and generation in protein structure modeling.
\end{abstract}

\section{Introduction}
Protein engineering aims to design and manipulate protein sequences and structures to achieve desired biological functions, and has become a central application of machine learning in the life sciences~\citep{notin2024machine}. 
Broadly, machine learning approaches to protein engineering can be categorized into two core tasks: protein understanding and protein generation. 
Protein understanding focuses on extracting functional and structural information from existing proteins, while protein generation aims to synthesize novel protein sequences or structures with desired properties~\citep{winnifrith2024generative, schmirler2024fine}.

Protein understanding models have achieved remarkable progress in recent years. 
Representative tasks include inverse folding~\citep{dauparas2022robust}, fold or structure prediction~\citep{jumper2021highly}, and functional annotation~\citep{ryu2019deep}, such as enzyme function prediction~\citep{yu2023enzyme}. 
Models such as protein language models~\citep{lin2023evolutionary} and protein structure encoders~\citep{zhang2022protein} have been shown to learn rich representations that are effective for downstream applications, including large-scale screening, functional annotation, and experimental validation. 
These understanding models are widely used as decision-making tools in protein engineering pipelines, where accurate representations are critical for selecting promising candidates.

In parallel, protein generative models have rapidly advanced, particularly with the adoption of diffusion-based and flow-based architectures for backbone and all-atom structure generation~\citep{watson2023novo, lu2025conditional, geffner2025proteina}. 
These models have demonstrated strong performance on challenging generation benchmarks, such as motif scaffolding and de novo structure design. 
However, despite their empirical success, an important question remains largely unexplored: do protein generative models learn representations that are useful for protein understanding tasks? 
In other words, beyond producing valid structures, do these models implicit learns semantic and functional information in a way that aligns with established protein understanding objectives?

\begin{figure*}
    \centering
    \includegraphics[width=0.8\linewidth]{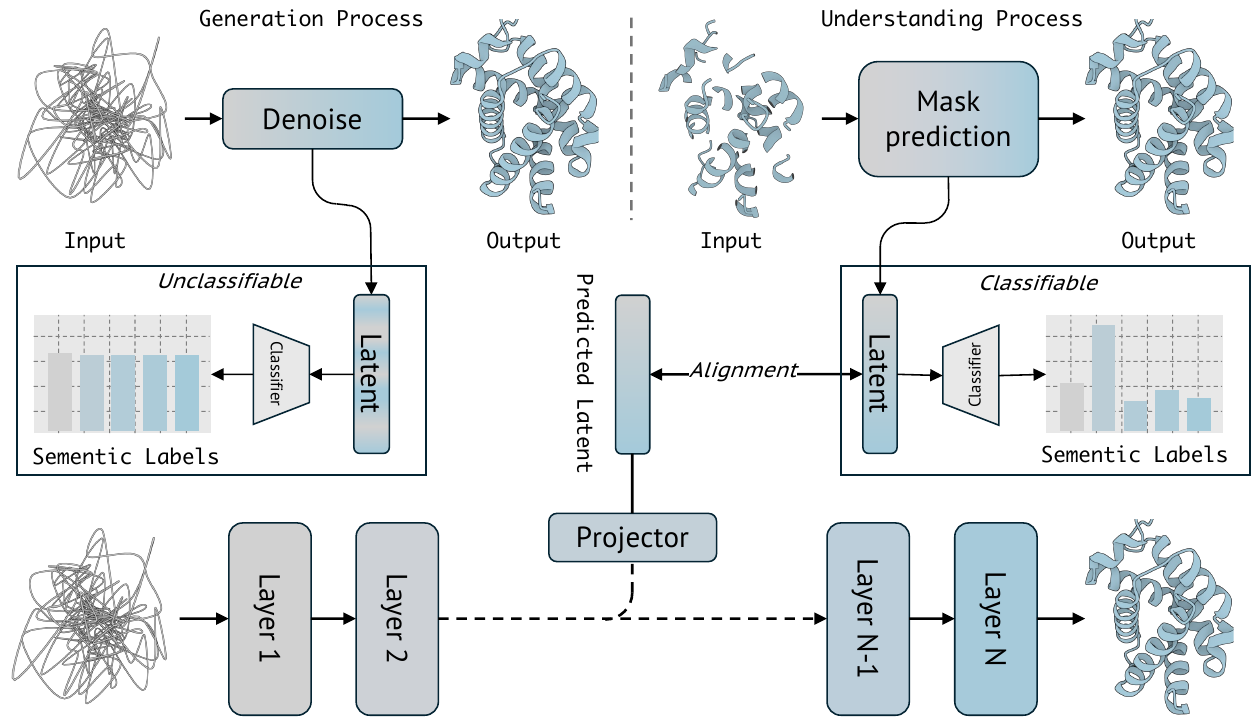}
    \caption{\textbf{Diagram of ReaPro-1c.} The Generation Process (left) employs a denoising diffusion model that maps noisy inputs to protein structures, yielding unclassifiable latent features with ambiguous semantic labels; the Understanding Process (right) utilizes a mask prediction model that reconstructs masked regions, producing classifiable latent representations with distinct semantic categories. To bridge this gap, we introduce a Projector within the intermediate layers of the generative model, aligning the denoising latents with the semantically rich understanding space.
    }
    \label{fig:main}
\end{figure*}
To answer this question, we conduct a systematic benchmarking study of state-of-the-art protein generative models on widely used protein understanding tasks. 
Our empirical results reveal a consistent gap: representations learned by generative models perform substantially worse than those from dedicated protein understanding models across multiple evaluation settings. 
This observation suggests that, similar to recent findings in vision~\citep{xiang2023denoising}, protein generative models may struggle to learn semantically meaningful representations when trained solely with generative objectives.

Motivated by this gap, we further explore whether protein generation can benefit from explicit representation-level guidance. 
Inspired by the REPresentation Alignment (REPA) framework~\citep{yu2024representation}, we propose to align the internal representations of protein diffusion models with pretrained protein understanding models during training. 
Experiments on the MotifBench~\citep{zheng2025motifbench} benchmark demonstrate that representation alignment significantly improves functional protein generation, increasing the MotifBench score of ProtPardelle-1c (cc58) from 39.2 to 47.1, corresponding to a 20\% relative improvement.
In another motif-conditional generation benchmark, the RFDiffusion benchmark, our methods overtakes baseline model on 22 out of 26 total motifs.
Notably, improvement was obtained by only half training steps with regard to the baseline model, which evidences the importance of representation alignment training.

\section{Methods}
\label{sec:method}
Our approach, ReaPro-1c, builds upon Protpardelle-1c~\citep{lu2025conditional}, a diffusion model for conditional protein structure generation. 
The key innovation is introducing representation alignment during training: we guide the generation model's intermediate representations to match those of a pretrained protein understanding model. 

\subsection{Model Architecture}

Protpardelle-1c employs a U-ViT architecture~\citep{bao2023all} that combines convolutional encoding/decoding with a central transformer. We briefly summarize its key components in the appendix~\ref{app:train}.




\textbf{Diffusion objective.} The model is trained with the EDM  framework~\citep{karras2022elucidating}. Given a noisy structure $\mathbf{x}_t$ at noise level $\sigma$, the model predicts the denoised structure $\hat{\mathbf{x}}$. The training loss uses preconditioning weights that depend on the noise level:

\begin{equation}
\mathcal{L}_{\text{struct}} = \mathbb{E}_{\sigma, \mathbf{x}_0, \mathbf{c}} \left[ w(\sigma) \cdot \left\| \hat{\mathbf{x}}(\mathbf{x}_\sigma, \sigma, \mathbf{c}) - \mathbf{x}_0 \right\|^2 \right]
\end{equation}

where $w(\sigma) = (\sigma^2 + \sigma_{\text{data}}^2) / (\sigma \cdot \sigma_{\text{data}})^2$ is the EDM loss weight, $\sigma_{\text{data}}$ is the data standard deviation, and $\mathbf{c}$ represents conditioning information (e.g., motif coordinates).


We introduce representation alignment at an intermediate layer of the U-ViT backbone. Specifically, at layer $k$ (default $k=5$) of the central transformer, we extract the hidden state $\mathbf{h}_k \in \mathbb{R}^{L \times d}$ and project it to the embedding space of a pretrained understanding model.

\textbf{Alignment head.} The projection is performed by an MLP with layer normalization and residual connection:

\begin{equation}
\mathbf{h}_{\text{align}} = \text{MLP}_{\text{align}}(\mathbf{h}_k) + \mathbf{W}_{\text{res}}(\mathbf{h}_k)
\end{equation}

where the $\text{MLP}:\mathbb{R}^{L \times d}\to \mathbb{R}^{L \times d_\text{align}}$ is a 3-layer network controlled by a hyperparameter $d_\text{proj}$.
The residual projection $\mathbf{W}_{\text{res}}: \mathbb{R}^d \to \mathbb{R}^{d_{\text{align}}}$ ensures dimensional compatibility between the diffusion model's hidden state $d$ and the understanding model's embedding dimension $d_\text{align}$.



\textbf{Target embedding.} The alignment target is computed by feeding the \emph{clean} ground-truth structure $\mathbf{x}_1$ into a frozen pretrained understanding model $f_{\text{understand}}$:

\begin{equation}
\mathbf{e}_{\text{target}} = f_{\text{understand}}(\mathbf{x}_1)
\end{equation}

We experiment with two types of understanding models: (1) ProteinMPNN/FA-MPNN~\citep{dauparas2022robust, widatalla2025sidechain}, a structure encoder trained for inverse folding; and (2) ESM2~\citep{lin2023evolutionary}, a protein language model trained on large-scale sequence data.

\subsection{Training Objective}

The total training loss combines the structure denoise objective with the alignment loss:

\begin{equation}
\mathcal{L}_{\text{total}} = \mathcal{L}_{\text{struct}} + \lambda \cdot \mathcal{L}_{\text{align}}
\label{eq:total_loss}
\end{equation}

where $\lambda$ is a hyperparameter controlling the alignment strength (default $\lambda=2.0$).




\textbf{Alignment loss.} $\mathcal{L}_{\text{align}}$ is the squared $\ell_2$ distance between the projected hidden state and the target embedding:

\begin{equation}
\mathcal{L}_{\text{align}} = \left\| \mathbf{h}_{\text{align}} - \mathbf{e}_{\text{target}} \right\|^2
\end{equation}

During inference, the alignment head is discarded, which guarantees no additional computational cost is incurred at test time.
We provide additional implementation details in Appendix~\ref{app:train}.

\section{Generative models do not learn discriminative features}
In this section, we presents a comprehensive evaluation of various protein models on tasks involving the classification of protein functions. 

\begin{table}[t] 
\caption{Performance comparison of different protein models using $F_{1}^{max}$ metric.}
\label{tab:protein-models}
\begin{center}
\resizebox{\textwidth}{!}{%
\begin{tabular}{lllccll}
\toprule
\textbf{Category} & \textbf{Models} & \textbf{Type/Description} & \textbf{EC} & \textbf{GO-mf} & \textbf{GO-bp} & \textbf{GO-cc} \\
\midrule
Tokenizers & La-Proteina VAE & All-Atom Continuous & 0.1420 & 0.1978 & 0.2723 & 0.2888 \\
           & Kanzi VQ-VAE    & Backbone Discrete   & 0.0901 & 0.1530 & 0.2489 & 0.3183 \\
\midrule
Generation & Protpardelle-1c (cc91) & All-Atom Diffusion & 0.1516 & 0.1946 & 0.2612 & 0.3380 \\
           & RFDiffusion            & Backbone Diffusion & 0.4242 & 0.2914 & 0.2753 & 0.3763 \\
\midrule
Understand & ESM         & Protein Language Model & \textbf{0.6376} & \textbf{0.6117} & \textbf{0.4317} & \textbf{0.4821} \\
           & ProteinMPNN & Structure Encoder      & 0.4528 & 0.4230 & 0.3375 & 0.4089 \\
\bottomrule
\end{tabular}%
}
\end{center}
\end{table}
\begin{figure*}[h]
    \centering
    \includegraphics[width=1\linewidth]{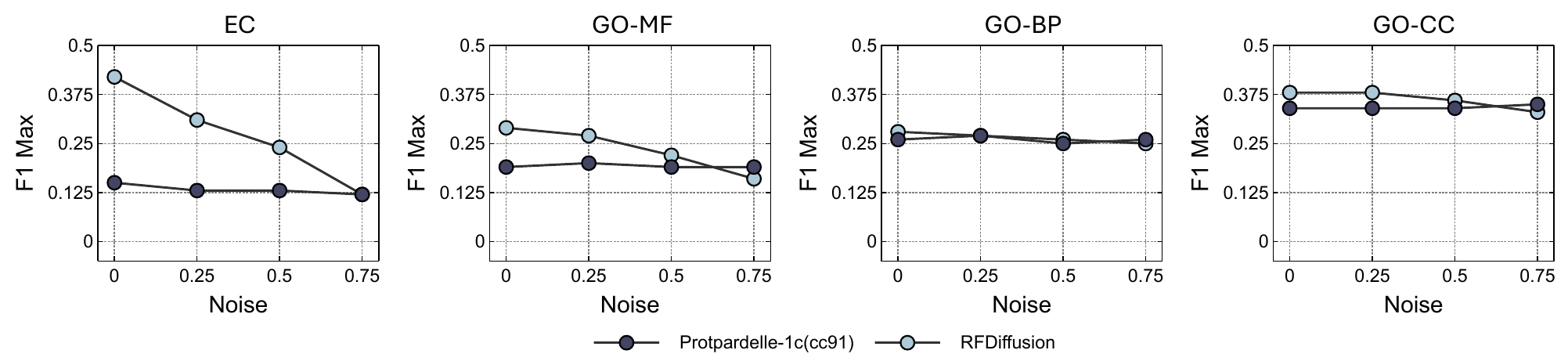}
    \caption{Functional classification performance under varying noise scales $\sigma \in [0, 0.25, 0.5, 0.75]$. Comparison of Protpardelle-1c and RFDiffusion on EC and GO tasks across varying noise levels.}
    \label{fig:und}
\end{figure*}
\subsection{Experiments Setup}
To assess the extent to which various models capture semantic biological knowledge, we evaluated their classification performance on Gene Ontology (GO) terms~\citep{ashburner2000gene} and Enzyme Commission (EC) numbers~\citep{bairoch2000enzyme}.  High performance on these benchmarks requires a model to internalize complex functional attributes that transcend the mere recognition of local structural geometries.
For comprehensive evaluation, we first selected two structural tokenizers, La-Proteina VAE~\citep{geffner2025proteina} (an all-Atom encoder that compresses side chains into a continuous vector) and Kanzi VQ-VAE~\citep{dilip2025flow} (a backbone encoder that encodes backbone atoms to a set of discrete tokens), which are capable of encoding and decoding protein geometries with very high precision.
To investigate whether the generative training objective facilitates the acquisition of discriminative biological features, we also evaluated two diffusion models, including the all-atom generative model Protpardelle-1c~\citep{lu2025conditional} (cc91) and the backbone generation model RFDiffusion~\citep{watson2023novo}. 
For Understanding models, we test the performance of the protein language model ESM~\citep{lin2023evolutionary} and the structure encoder ProteinMPNN~\citep{dauparas2022robust} as baselines. 
We provide more details in Appendix~\ref{app:cla}.

\subsection{Results}
Table~\ref{tab:protein-models} summarizes the classification performance of various protein models across the EC and GO benchmarks, as quantified by the $F1_{max}$ metric.  
Consistent with observations in computer vision, understanding models (ESM and ProteinMPNN) significantly outperform both structure tokenizers and diffusion models in all classification tasks.
These results suggest that while tokenizers are capable of accurately reconstructing protein structures from a latent space, they fail to internalize high-level functional semantics during the bottleneck compression and reconstruction process. 
Similarly, our findings demonstrate that the internal representation of diffusion models remains inferior to that of specialized protein encoders. 
This gap indicates that while generative objectives excel at structural modeling, they may not naturally align with the capture of discriminative biological function without further refinement.

To investigate whether the inferior performance stems from the noisy training distribution inherent in diffusion processes, we analyze the classification performance of RFDiffusion and Protpardelle-cc91 across varying noise levels $\sigma \in [0, 0.25, 0.5, 0.75]$. 
As illustrated in Fig~\ref{fig:und}, functional discriminability for RFDiffusion exhibits a pronounced downward trend as noise intensity increases, whereas this sensitivity is notably weaker for Protpardelle-cc91. In Enzyme Commission (EC) number prediction, RFDiffusion significantly outperforms Protpardelle-cc91 at lower noise levels ($\sigma \leq 0.50$). 
However, it suffers from a sharp decay in performance as noise increases, with the $F^{max}$ score plummeting from $0.4242$ at the zero-noise limit ($\sigma=0$) to $0.1193$ at $\sigma=0.75$. 
This initial superiority of RFDiffusion stems from the high-fidelity structural priors embedded within the RoseTTAFold architecture, reflecting the paradigm that structure dictates function. 
Nevertheless, as noise intensity scales, these structural priors become corrupted, leading to the observed degradation. 

\section{Understanding model helps unconditional generation.}
\begin{figure*}[t]
    \centering
    \includegraphics[width=0.9\linewidth]{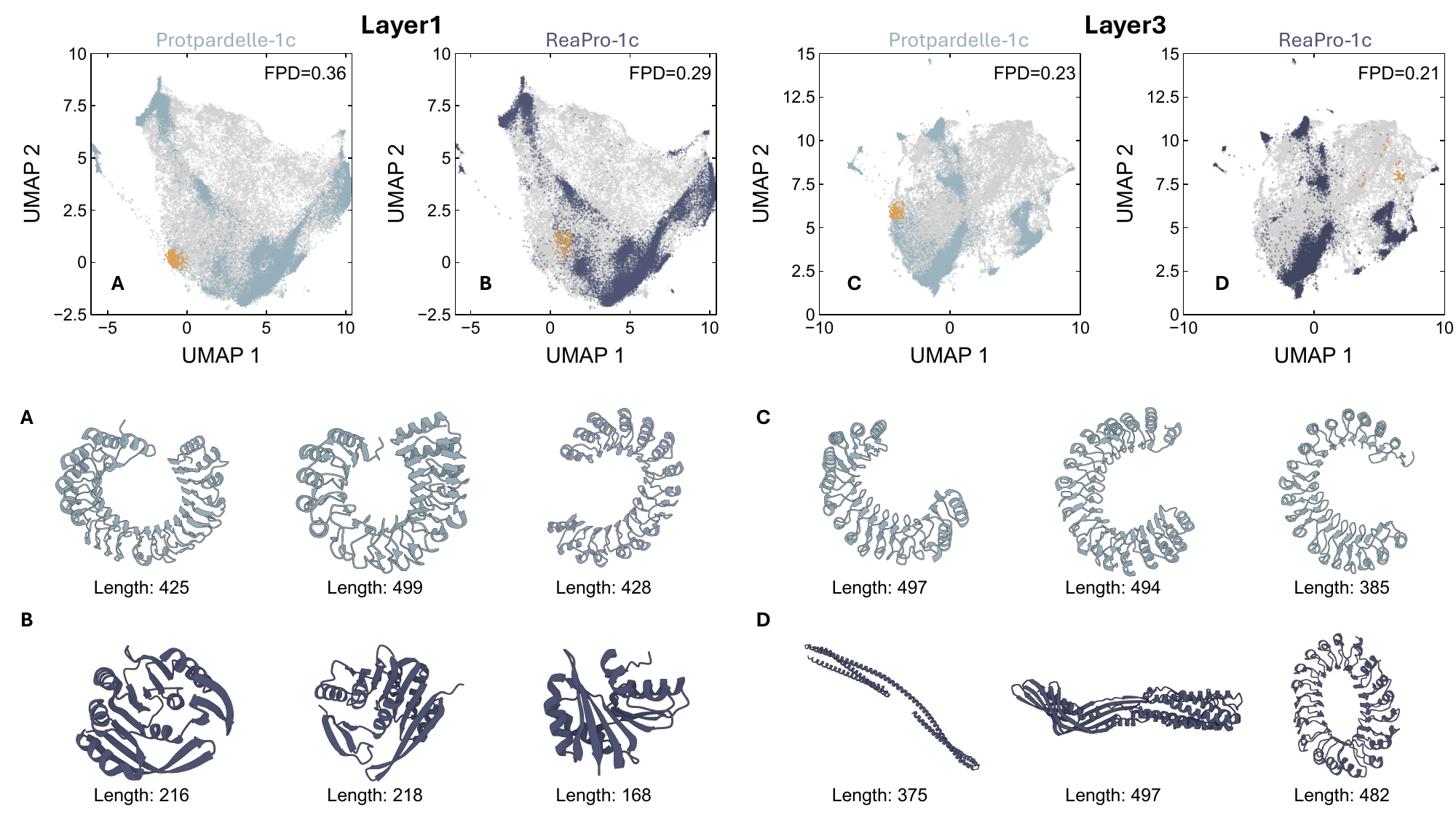}
    \caption{\textbf{Distribution coverage comparison between Protpardelle-1c and ReaPro-1c.} UMAP visualization of ProteinMPNN embeddings at Layer 1 and Layer 3. {\color{gray}Gray} points represent CATH 4.4-S40 reference structures; {\color{blue}blue} points represent generated samples. {\color{orange}Orange} points highlight regions where one model shows higher sampling density than the other. FPD scores indicate that ReaPro-1c achieves better alignment with the reference distribution at both layers.
    (\textbf{A}) Visualization of structures where Protpardelle-1c shows higher sampling density than ReaPro-1c; (\textbf{B}) Visualization structures where Protpardelle-1c shows higher sampling density than ReaPro-1c.
    (\textbf{C, D}) Additional examples from Layer 3 analysis.
    }
    \label{fig:uncondition}
\end{figure*}
In this section, we evaluate the performance of ReaPro-1c on unconditional protein structure generation. 
\subsection{Experiment Setup}
We use CATH 4.4-S40~\citep{sillitoe2021cath} as the reference dataset to evaluate the distributional coverage of generated structures. 
This dataset contains 601,328 protein domains with less than 40\% sequence identity, representing a diverse set of naturally occurring protein structures.
To ensure a fair comparison, we adopt a length-matched sampling strategy. 
Specifically, we exclude proteins with lengths over 500 and compute the length distribution of proteins in CATH 4.4-S40, resulting in 28,449 structures.
For each length $L$ that appears $N_L$ times in the reference set, we generate $N_L$ samples of length $L$ from both the baseline model and our model. 
This approach ensures that any observed differences in distribution are attributable to the models' generation capabilities rather than length bias. 
The sampling hyperparameters are kept identical for both models (see Appendix~\ref{app:eval} for details).

We adopt the SHAPES framework~\citep{lu2025assessing} to assess the coverage of protein structure space. 
SHAPES quantifies distributional similarity using Fr\'echet Protein Distance (FPD), analogous to Fr\'echet Inception Distance (FID) in image generation.
Specifically, we extract structural embeddings using ProteinMPNN at different layers (Layer 1 and Layer 3) and compute the FPD between the generated samples and the reference CATH structures. 
Lower FPD values indicate better alignment with the natural protein structure distribution.

\subsection{Results}
Fig.~\ref{fig:uncondition} presents the UMAP visualization of ProteinMPNN embeddings for both models alongside the CATH reference set. 
The quantitative FPD scores reveal that ReaPro-1c consistently achieves better distributional alignment with natural protein structures.
The improvement is more pronounced at Layer 1 (0.29 vs 0.36), which captures local structural features, suggesting that the understanding model guidance particularly enhances the generation of chemically plausible local geometries. 
At Layer 3, which encodes more global structural information, both models perform comparably (0.21 vs 0.23), with ReaPro-1c maintaining a slight advantage.

To gain deeper insights into the differences between the two models, we randomly select regions in the embedding space where one model exhibits significantly higher sampling density than the other (Orange dots in Fig.~\ref{fig:uncondition}). 
We then extract the structures falling into these differentially sampled regions and cluster them using Foldseek~\citep{van2024fast}. The cluster centers are visualized to characterize the structural motifs that distinguish each model's generation behavior.
As shown in Fig.~\ref{fig:uncondition}A, regions where Protpardelle-1c exhibits higher sampling density are dominated by large $\alpha$-helical bundles and ring-like architectures (e.g., lengths 425, 499, 428). 
This suggests that the baseline model may have a bias toward generating highly regular, designable structures with prominent secondary structure content.
In contrast, Fig.~\ref{fig:uncondition}B shows that regions where ReaPro-1c exhibits higher sampling density feature more diverse structural classes, including $\beta$-sheet rich folds (lengths 216, 218) and mixed $\alpha$/$\beta$ structures (length 168). 
The Layer 3 analysis (Fig.~\ref{fig:uncondition}D) further reveals that ReaPro-1c generates more extended and irregular structures, including elongated helical assemblies (length 375) and complex multi-domain arrangements (lengths 497, 482).

\section{Understanding model helps conditional generation.}
\label{sec:conditional}
\begin{figure*}
    \centering
    \includegraphics[width=1\linewidth]{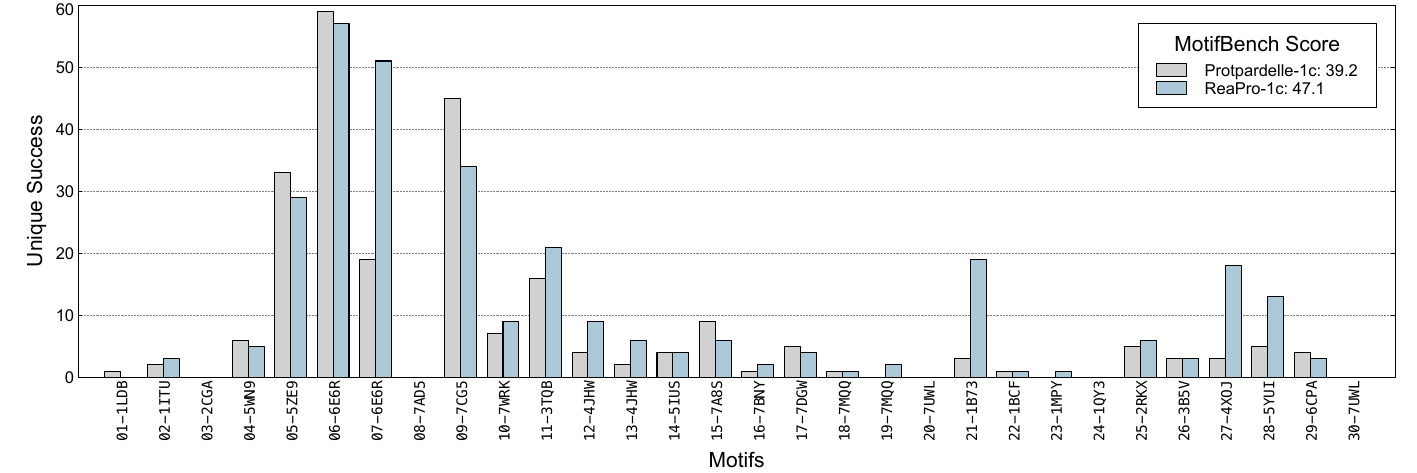}
    \caption{MotifBench results comparing Protpardelle-1c and ReaPro-1c. The bar plot shows the number of unique successful designs per motif. ReaPro-1c achieves a total score of 45.6 compared to Protpardelle-1c's 39.2, successfully designing 25 out of 30 motifs versus 23 for the baseline.}
    \label{fig:motifbench}
\end{figure*}
\begin{figure*}
    \centering
    \includegraphics[width=1\linewidth]{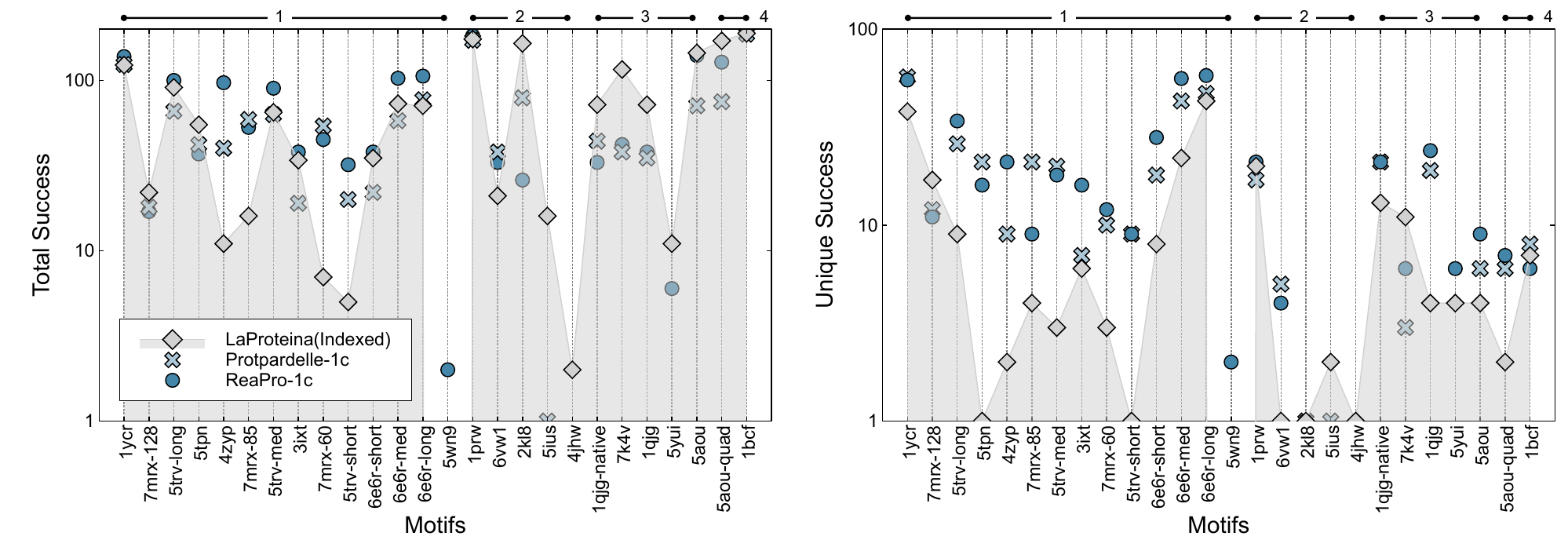}
    
    
    \includegraphics[width=1\linewidth]{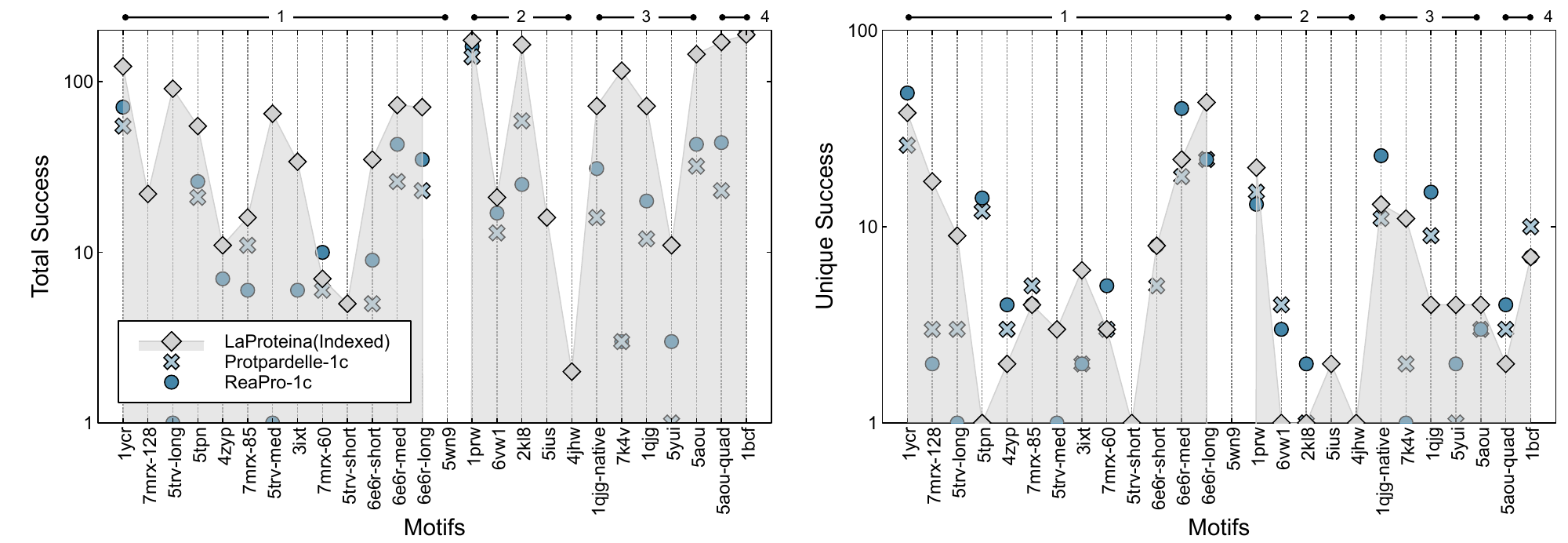}
    \caption{\textbf{Top: RFDiffusion backbone generation benchmark results.} (\textbf{Left}) Total success count per motif. (\textbf{Right}) Unique success count after clustering. Top annotations indicate the number of motif segments (1--4). \textbf{Bottom: RFDiffusion all-atom generation results.} (\textbf{Left}) Total success count. (\textbf{Right}) Unique success count.}
    \label{fig:rfdiffusion-both}
\end{figure*}
In this section, we evaluate the effectiveness of our approach on conditional protein generation tasks, specifically focusing on motif scaffolding. 
Motif scaffolding is a protein design task where the goal is to generate a protein structure that preserves a given functional motif (a set of residues with specific 3D coordinates) while designing the surrounding scaffold. 
This task is critical for designing proteins with desired functional properties, such as binding sites or catalytic centers.

\subsection{Motifbench}
We evaluate on MotifBench~\citep{zheng2025motifbench}, a standardized benchmark containing 30 diverse motif-scaffolding problems. 
For each target motif, we generate 100 structures using the same sampling hyperparameters as the baseline.
Following the evaluation protocol of Protpardelle-1c~\citep{lu2025conditional}, we use ProteinMPNN \citep{dauparas2022robust} to design 8 amino acid sequences for each generated backbone, then fold these sequences with ESMFold~\citep{lin2023evolutionary}. 
A design is considered successful if the folded N, C$_\alpha$, and C of motif structure achieves an RMSD of less than 1\AA\ to the target motif, and the C$_\alpha$ of scaffold achieves an RMSD of less than 2\AA.

As shown in Figure~\ref{fig:motifbench}, ReaPro-1c achieves substantial improvements over the baseline. 
The motifbench score increases from 39.2 to 47.1, representing a 20\% improvement. 
More importantly, ReaPro-1c successfully designs scaffolds for 25 out of 30 motifs, compared to 23 for Protpardelle-1c. 
This demonstrates that incorporating understanding model guidance not only increases the diversity of successful designs but also expands the range of motifs that can be designed.

\subsection{RFDiffusion Benchmark}

We further evaluate on the RFDiffusion motif benchmark~\citep{watson2023novo}, which contains 26 challenging motif-scaffolding problems with varying complexity (1--4 motif segments). 
We compare against both Protpardelle-1c and La-Proteina~\citep{geffner2025proteina}, a state-of-the-art flow matching approach for joint sequence-structure generation.
For each of the 26 motifs, we generate 200 backbone structures and design a single sequence per structure using ProteinMPNN.
Success is determined by ESMFold folding with motif N, C$_\alpha$, and C RMSD less than 1\AA, and the C$_\alpha$ of the scaffold achieves an RMSD of less than 2\AA.
We report both the total success count (all successful designs) and unique success count (clustered at TM-score $<$ 0.5 using Foldseek~\citep{van2024fast}).

Figure~\ref{fig:rfdiffusion-both} presents the comparison results. 
ReaPro-1c demonstrates strong performance compared with Protpardelle-1c and La-Proteina.
For total success, ReaPro-1c matches or exceeds La-Proteina on 15 out of 26 motifs, and outperforms Protpardelle-1c on 22 out of 26 motifs.
The advantage becomes more pronounced when measuring structural diversity. 
ReaPro-1c matches or exceeds La-Proteina on 21 out of 26 motifs, and outperforms Protpardelle-1c on 17 out of 26 motifs.
This result suggests that the understanding model guidance enables ReaPro-1c to generate a more diverse set of successful scaffolds, even when the total success counts are comparable. 
The improvement is particularly notable for multi-segment motifs (2--4 segments), where the structural constraints are more complex.

\subsection{All-Atom Generation}

To evaluate the extension of our approach to all-atom generation, we train a variant following the CC91 setting of protpardelle-1c \citep{lu2025conditional}. 
This model generates full atomistic structures, including side chains. 
We use the same evaluation protocol as Protpardelle-1c, sample 200 structures per motif, and redesign a single sequence per structure using ProteinMPNN.
\begin{figure*}
\centering
\includegraphics[width=0.9\linewidth]{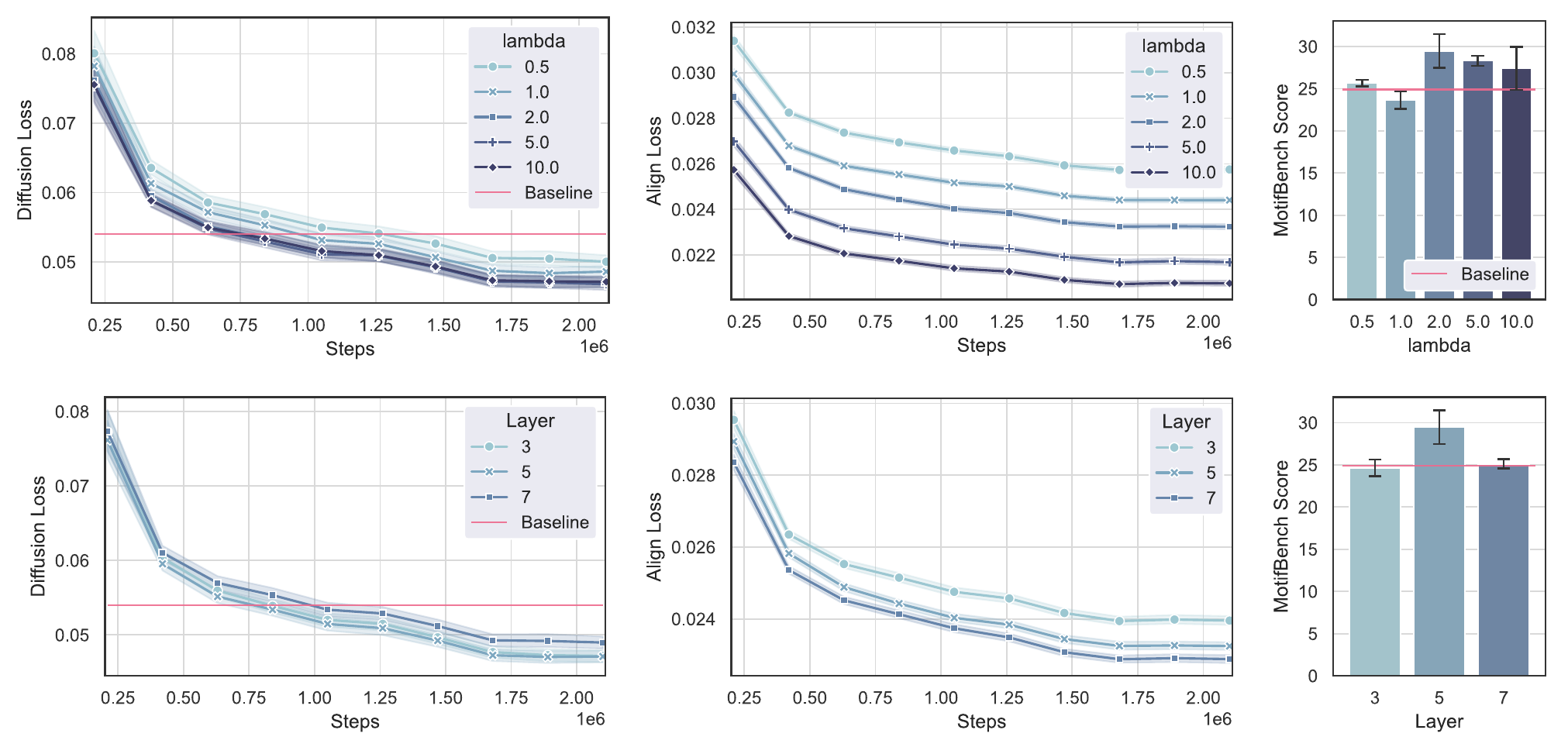}
\caption{\textbf{Ablation on alignment loss weight ($\lambda$) and alignment layers.} (\textbf{Left}) Diffusion loss curves during training. (\textbf{Middle}) Alignment loss curves. (\textbf{Right}) MotifBench scores for different $\lambda$ and layers.}
\label{fig:ablate_ll}
\vspace{-1em}
\end{figure*}

As shown in Figure~\ref{fig:rfdiffusion-both}, the all-atom generation task presents greater challenges.
In total success, ReaPro-1c matches or exceeds La-Proteina on 6 out of 26 motifs, but significantly outperforms Protpardelle-1c on 23 out of 26 motifs.
In unique success, ReaPro-1c matches or exceeds La-Proteina on 12 out of 26 motifs, and outperforms Protpardelle-1c on 19 out of 26 motifs. 
While there remains a performance gap to La-Proteina (which employs a specialized partially latent flow matching framework for all-atom generation), ReaPro-1c substantially narrows the gap between Protpardelle-1c and La-Proteina. 
This demonstrates that the understanding model guidance transfers effectively to the more complex all-atom generation setting, even without architectural modifications specifically tailored for atomistic modeling.

\section{Ablation Studies}
\label{sec:ablation}
\begin{wrapfigure}{r}{0.5\textwidth} 
\centering
\vspace{-3em}  
\includegraphics[width=\linewidth]{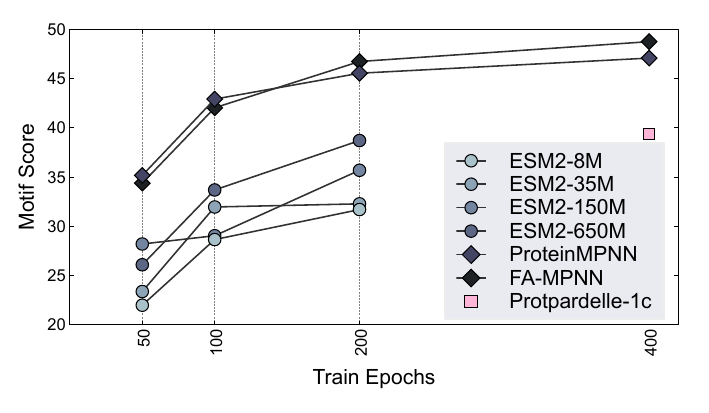}
\caption{\textbf{Ablation on understanding model choice.} MotifBench scores of models trained with different understanding models for representation alignment. All models are trained for 200 epochs with identical hyperparameters except for the alignment target. The baseline (Protpardelle-1c) is trained for 416 epochs.}
\label{fig:ablate_esmmpnn}
\vspace{-3em}  
\end{wrapfigure}
In this section, we conduct ablation studies to investigate the key design choices in our approach: (1) the choice of understanding model for alignment, (2) the weight of the alignment loss, and (3) the layer at which alignment is performed. 
All ablations are conducted on the MotifBench benchmark 
following the evaluation protocol described in Section~\ref{sec:conditional}.

\subsection{Choice of Understanding Model}
We first investigate which understanding model provides the most effective guidance for generation. We compare ESM2 models of varying scales (8M, 35M, 150M, 650M parameters)~\citep{lin2023evolutionary} against ProteinMPNN~\citep{dauparas2022robust}, a structure encoder specifically designed for protein inverse folding.

As shown in Figure~\ref{fig:ablate_esmmpnn}, we observe a clear scaling trend with ESM2 models: larger models consistently yield better generation performance. ESM2-650M achieves a MotifBench score of 38.7 at 200 epochs, approaching the baseline performance (39.4) despite the baseline being trained for 416 epochs. 
This demonstrates that alignment with stronger sequence understanding models leads to faster convergence and better final performance.

Notably, ProteinMPNN achieves the highest score (47.1), significantly outperforming all ESM2 variants. 
We hypothesize that while ESM2 excels at protein understanding tasks (e.g., fitness prediction, contact prediction), there exists an inherent modality gap between sequence representations and 3D structures. 
ProteinMPNN bridges this gap by operating directly on atomic coordinates.
To verify this, we conduct an experiments that employ a different structure encoder FA-MPNN~\citep{widatalla2025sidechain} to check whether this trend persists.
The result of FA-MPNN suggest that structure-aware understanding models provide more effective guidance for structure generation, even when sequence-based models like ESM2 are scaled up.

\subsection{Alignment Loss Weight}

Next, we examine the effect of the alignment loss weight $\lambda$ in Equation~\ref{eq:total_loss}. We vary $\lambda \in \{0.5, 1.0, 2.0, 5.0, 10.0\}$ while keeping all other hyperparameters fixed. 
For the motif benchmark, we report the motif score of 3 successive epochs of each setting: 199, 200, and 201, and we design 1 sequence for each structure.

Figure~\ref{fig:ablate_ll} (upper) shows the diffusion loss and alignment loss curves during training. 
Higher $\lambda$ values lead to lower alignment loss, indicating a stronger alignment with the understanding model. Notably, stronger alignment also shows lower diffusion loss, evidence that alignment with an understanding model can facilitate generative ability.
Over-aggressive alignment makes the improvement of diffusion loss marginal; this trend also coincides with the motifscore, where the performance starts to decline for $\lambda > 2.0$.
This suggests an optimal balance: insufficient alignment fails to provide meaningful guidance, while excessive alignment may constrain the generator's exploration of the structure space. 
We use $\lambda = 2.0$ for all main experiments.

\subsection{Alignment Layer Selection}

Finally, we investigate at which layer of the understanding model the alignment should be performed. We test aligning at layers 3, 5, and 7 of our model.

As shown in Figure~\ref{fig:ablate_ll} (lower), aligning at layer 5 achieves the best MotifBench performance.
Layer 3 (early features) and layer 7 (late features) both underperform, suggesting that mid-level representations strike the best balance between capturing local structural details and global architectural patterns. 
The alignment loss curves (middle panel) show that layer 7 achieves the lowest alignment loss, likely because late-layer features are more compressed and easier to match. 
However, this does not translate to better generation performance, indicating that aligning to overly abstract representations may lose structural details critical for generation. 
We align at layer 5 for all main experiments.

\section{Discussion}
\label{sec:discussion}

In this work, we reveal a consistent gap between protein understanding and generation models: state-of-the-art generative models learn suboptimal representations for discriminative tasks. 
To bridge this gap, we propose ReaPro-1c, which aligns protein diffusion models with pretrained understanding models via a lightweight projection head. 
Experiments demonstrate that this simple alignment yields substantial improvements across benchmarks---boosting MotifBench scores by 20\%, improving distribution coverage on CATH, and accelerating convergence. 
Our ablations further reveal that structure encoders (ProteinMPNN) provide more effective guidance than sequence encoders (ESM2), highlighting the importance of matching modalities between the understanding target and the generation task.

\textbf{Limitations and future work.} 
First, we currently align at a single layer; multi-layer alignment could provide richer guidance. 
Second, extending representation alignment to other protein design tasks, such as binder design and enzyme engineering, and exploring alignment with multiple complementary understanding models are promising avenues for future research.
More broadly, recent studies on unifying understanding and generation models have achieved impressive results (RAE~\citep{zheng2025diffusion}), suggesting the potential of directly generating proteins from semantically rich representations, such as those learned by AlphaFold~\citep{jumper2021highly}.

\bibliography{reference}
\bibliographystyle{plain}


\clearpage
\appendix
\section{Preliminaries and Related Works}
\label{sec:rw}

\subsection{Protein Understanding}
\label{sec:rw_und}
\textbf{Protein Structure}
determines how protein performs their biological function~\citep{huang2016coming, notin2024machine}.
In computational settings, protein structures are commonly represented as sets of 3D atomic coordinates.
In backbone-centric settings, we model the rigid frames or coordinates of the backbone atoms (N, C$_\alpha$, C, O). 
Formally, the backbone coordinates are denoted as $\mathbf{X}^{\text{bb}} \in \mathbb{R}^{L \times 4 \times 3}$, where $L$ is the sequence length.
For more detailed modeling, a common standardized representation is \emph{Atom37}, which stores up to 37 atom positions per residue: $\mathbf{X}^{37} \in \mathbb{R}^{L \times 37 \times 3}$ with a residue-dependent mask selecting the valid subset~\citep{watson2023novo}.

  
To learn from structure, proteins are frequently represented as graphs
$G=(V,E)$ where nodes correspond to residues (or atoms) and edges encode spatial proximity or chemical bonds~\citep {hermosilla2022contrastive}.
Representative methods like GearNet~\citep{zhang2022protein}, utilize graph contrastive learning~\citep{you2020graph, qiu2020gcc} to learn from two augmented structures.
ATOMICA~\citep{fang2025atomica} learns complex representations by recovering the original atom coordinates and types from a corrupted input.
However, due to the scarcity of protein structures, these methods are less competitive than large-scale pre-trained sequence models~\citep{jamasb2024evaluating}.

\textbf{Protein Sequence}
are represented as strings of amino acid tokens from a vocabulary of size 20 (standard amino acids) plus special tokens. Formally, a sequence is denoted as $\mathbf{s} = (s_1, s_2, \ldots, s_L)$ where $s_i \in \mathcal{V}$ and $\mathcal{V}$ is the amino acid vocabulary.

Benefiting from the large-scale protein sequence data, protein language models (PLMs) have become the dominant approach for protein understanding. 
ESM2~\citep{lin2023evolutionary} demonstrates that scaling model size and training data leads to emergent capabilities in structure prediction and functional understanding.
Recently, there has been growing interest in integrating structural information into PLMs. 
SaProt~\citep{su2023saprot} and Prot-T5~\citep{heinzinger2024sty} introduce a structure-aware vocabulary~\citep{van2022foldseek} that encodes both sequence and local structural context. 
This structure-aware vocabulary is small and can only reconstruct rough structures.
To eliminate the potential reconstruction bottleneck of structure vocabulary, ESM3~\citep{hayes2025simulating}, DPLM2~\citep{wang2024dplm} using a VQ-VAE~\citep{van2017neural}, served as a structure tokenizer to first compress structures into a discrete token, then joint training with sequence data~\citep{medvedev2025biotoken}.

\subsection{Protein Generation}
\label{sec:rw_gen}
\textbf{From backbone to inverse-folding.}
As different amino acids have different atoms, a popular design pipeline is using the diffusion model to generate the backbone atoms (N, C, Ca, O) first, including RFDiffusion~\citep{watson2023novo}, Protpardelle~\cite{lu2025conditional}, etc.~\citep{yim2023se, ingraham2023illuminating,yim2023fast,bose2023se,huguet2024sequence}, or Ca-only methods (Proteina and Genie 2~\citep{geffner2025proteina, lin2024out}). 
Followed by an inverse folding method~\citep{dauparas2022robust,qiu2024instructplm} to predict the label of the side chain.
Similar to the understanding model, another line of work adopts a structure tokenizer to compress protein structures into a discrete token first, then trains a GPT-style auto-regressive model to generate novel protein structures~\citep{lu2024structure,gaujac2024learning,gao2025foldtoken,dilip2025flow}.

\textbf{Co-design.}
Recent work increasingly emphasizes co-design, i.e., jointly generating
sequence and structure, potentially at the all-atom level.
La-Proteina~\citep{geffner2025proteina} proposes a partially latent flow matching framework for joint generation of protein sequence and fully atomistic structure: it maintains explicit backbone variables while encoding per-residue sequence and side-chain details into fixed-size continuous latents.
This hybrid design aims to avoid mixed continuous--categorical modeling complexity and improve scalability for long proteins.

\subsection{Alignment of Understanding and Generation}
\label{sec:rw_align}
A recurring observation across domains is that generative denoising objectives
can induce discriminative representations~\citep{li2023your, chen2024deconstructing, assran2023self}, yet these representations often lag behind those learned by specialized self-supervised encoders.
This motivates approaches that explicitly inject or align semantic representations to improve generative modeling.
Representation Alignment (REPA) proposes a simple regularization that aligns hidden states of a denoising network operating on noisy inputs with representations computed by an external pretrained encoder on the corresponding clean inputs~\citep{yu2024representation, leng2025repa}.
Instead of adding an auxiliary alignment loss, representation autoencoders(RAEs) replace the conventional VAE encoder in latent diffusion with a frozen pretrained representation encoder (e.g., DINO/SigLIP/MAE) paired with a lightweight decoder, forming a latent space that is both reconstruction-capable and semantically rich~\citep{zheng2025diffusion}.
This direction argues that pretrained semantic representations can directly serve as effective latents for diffusion transformers and can yield faster convergence and improved generation quality~\citep{tong2026scaling}.

\section{Implement Details of Classification Tasks}
\label{app:cla}
\subsection{la-proteina}
The implementation utilized the publicly available codebase~\url{https://github.com/NVIDIA-Digital-Bio/la-proteina}.
For the input pdb, we perform \texttt{CenterStructureTransform} and \texttt{ChainBreakPerResidueTransform}.
Result $L \times 16$ embeddings.



\subsection{Kanzi}
The implementation utilized the publicly available publicly-released codebase~\url{https://github.com/rdilip/kanzi}.
Only support 256, we extend the max length to 1024.
For each input pdb, we use the discrete embeddings corresponding to the encoded cod
Result $L \times 256$ embeddings.

\subsection{ProtPardelle-1c}
The implementation utilized the publicly available codebase~\url{https://github.com/ProteinDesignLab/protpardelle-1c/}. Specifically, varying levels of noise ($0.0, 0.25, 0.50,$ and $0.75$) were added to the protein structures to extract corresponding protein embeddings across these different noise regimes. We extracted the hidden state representations from the U-ViT module, yielding feature dimensions of $L \times 256$.

\subsection{RFDiffusion}
The implementation utilized the publicly available codebase~\url{https://github.com/RosettaCommons/RFdiffusion}. 
During the forward pass of the diffusion process, the state\_prev embeddings were extracted across four noise regimes ($0.0, 0.25, 0.50,$ and $0.75$). 
This process resulted in feature representations with dimensions of $L \times 64$.

\subsection{ESM2}
For the sequence-based baseline, we utilized the ESM-2 650M parameter model (\texttt{esm2\_t33\_650M\_UR50D}). Protein representations were obtained by extracting the embeddings from the $33^{rd}$ layer (the final hidden layer), resulting $L \times 1280$ embeddings.

\subsection{ProteinMPNN}
The implementation followed the standard ProteinMPNN architecture for structure-based encoding. We extracted the output embeddings directly from the structural encoder to serve as the fixed-length representation for the downstream functional classification tasks. Result $L \times 128$ embeddings.

\subsection{Classification head}
To aggregate the residue-level embeddings into a fixed-size protein-level representation, we employ an attention-based pooling mechanism. Let $\mathbf{H} = [\mathbf{h}_1, \mathbf{h}_2, \dots, \mathbf{h}_L] \in \mathbb{R}^{L \times D}$ denote the sequence of embeddings for a protein of length $L$. The attention score $e_i$ for the $i$-th residue is computed as:$$e_i = \mathbf{w}^\top \tanh(\mathbf{W}_a \mathbf{h}_i + \mathbf{b}_a) + b_w$$where $\mathbf{W}_a \in \mathbb{R}^{H \times D}$ and $\mathbf{w} \in \mathbb{R}^{H}$ are learnable parameters of the hidden and scoring layers, respectively. To ensure the model focuses only on valid residues, we apply a mask $M$ such that $e_i = -\infty$ for padding tokens. The normalized attention weights $\alpha_i$ are then calculated via the Softmax function:$$\alpha_i = \frac{\exp(e_i)}{\sum_{j=1}^{L} \exp(e_j)}$$The final aggregated protein representation $\mathbf{z} \in \mathbb{R}^D$ is obtained as the weighted sum:$$\mathbf{z} = \sum_{i=1}^{L} \alpha_i \mathbf{h}_i$$The global vector $\mathbf{z}$ is subsequently passed through a readout network $\Phi$ to generate the final predictions $\hat{\mathbf{y}}$:$$\hat{\mathbf{y}} = \mathbf{W}_2 \cdot \text{Dropout}(\text{ReLU}(\mathbf{W}_1 \mathbf{z} + \mathbf{b}_1)) + \mathbf{b}_2$$where $\mathbf{W}_1$ and $\mathbf{W}_2$ represent the weight matrices of the two-layer Multi-Layer Perceptron (MLP). This architecture allows the model to adaptively highlight functional motifs within the protein sequence for classification.

\section{Training Details of ReaPro-1c}
\label{app:train}
\subsection{Model Architecture}
We adopt the same model architecture as the original Protpardelle-1c~\citep{lu2025conditional}.

\textbf{Input representation.} The model operates on all-atom protein coordinates $\mathbf{X} \in \mathbb{R}^{L \times A \times 3}$, where $L$ is the sequence length and $A$ is the number of atoms per residue (37 for Atom37 representation).

\textbf{U-ViT backbone.} The architecture consists of: (1) downsampling convolutional blocks that process atomic coordinates into patch embeddings; (2) a central Time-Conditional Transformer with rotary positional embeddings; and (3) upsampling convolutional blocks that reconstruct the full-resolution output. 
The transformer contains 10 layers by default, with each layer comprising time-conditional self-attention and feed-forward modules.

We use cc58 and cc91 as the base models for backbone design and all-atom design, respectively.
For the alignment head, we use a 3-layer MLP to project the original hidden state $\mathbf{h}_{k} \in \mathbb{R}^{L \times d}$ into the space of the understanding hidden state $\mathbf{h}_{\text{align}}$. 
The pseudo code of the alignment head is shown in Algorithm~\ref{alg:alignment_head}.

\begin{algorithm}[h]
\caption{Alignment Head Architecture}
\label{alg:alignment_head}
\begin{verbatim}
class AlignmentHead(nn.Module):
    def __init__(self, hidden_size, projector_dim, z_dim):
        super().__init__()
        # Residual projection for dimensional matching
        self.residue_proj = nn.Linear(hidden_size, z_dim)
        
        # 3-layer MLP with SiLU activation
        self.mlp = nn.Sequential(
            nn.LayerNorm(hidden_size),
            nn.Linear(hidden_size, projector_dim),
            nn.SiLU(),
            nn.Linear(projector_dim, projector_dim),
            nn.SiLU(),
            nn.Linear(projector_dim, z_dim),
            nn.LayerNorm(z_dim)
        )
        
    def forward(self, h_k):
        # MLP pathway
        out = self.mlp(h_k)
        
        # Residual connection with projection
        residue = self.residue_proj(h_k)
        
        return out + residue
\end{verbatim}
\end{algorithm}

Tab.\ref{tab:model_param} shows the detailed model parameters we used in the main results.
\begin{table}[h]
    \centering
    \begin{tabular}{l|r}
    \toprule
    Name & Value \\
    \midrule
    layers & 10 \\
    n\_channel & 256 \\
    dim\_head & 32 \\
    rope\_dim & 32 \\
    n\_blocks\_per\_layer & 2 \\
    n\_heads & 8 \\
    patch\_size & 1 \\
    \bottomrule
    \end{tabular}
    \caption{Model parameters of ReaPro-1c.}
    \label{tab:model_param}
\end{table}

\subsection{Dataset}
Following the same setting of Protpardelle-1c, we adopt the augmented Ingraham CATH dataset for training.
The dataset includes 337,936 structures.

\subsection{Training Parameters}
We use Adam~\citep{kingma2014adam} optimizer with a batch size of 32.
We start the training process with a 1000-step linear warm-up, increase the learning rate to $1\times 10^{-4}$, and using an cosine annealing to decrease the learning rate to $1\times 10^{-6}$ in 2,000,000 steps.
We use the same crop config as Protpardelle-1c.
We train the model for 401 epochs.
For the all-atom model, we use the same setting but train for 201 epochs.
All experiments are done on a single Nvidia A100 GPU.

\section{Evaluation Details of ReaPro-1c}
\label{app:eval}
\subsection{Motifbench}
To get the final score, we first generate 100 structures and design 1 sequence per structure with ProteinMPNN.
We use models of 399, 400, and 401 epochs.
Then we choose models with the best score and generate 100 structure deign 8 sequences per structure to get the final results.
We use the same sampling parameters as original Protpardelle-1c, with \textsc{schurns = 200} and \textsc{step\_scales = 1.2}.
\begin{figure}
    \centering
    \includegraphics[width=1\linewidth]{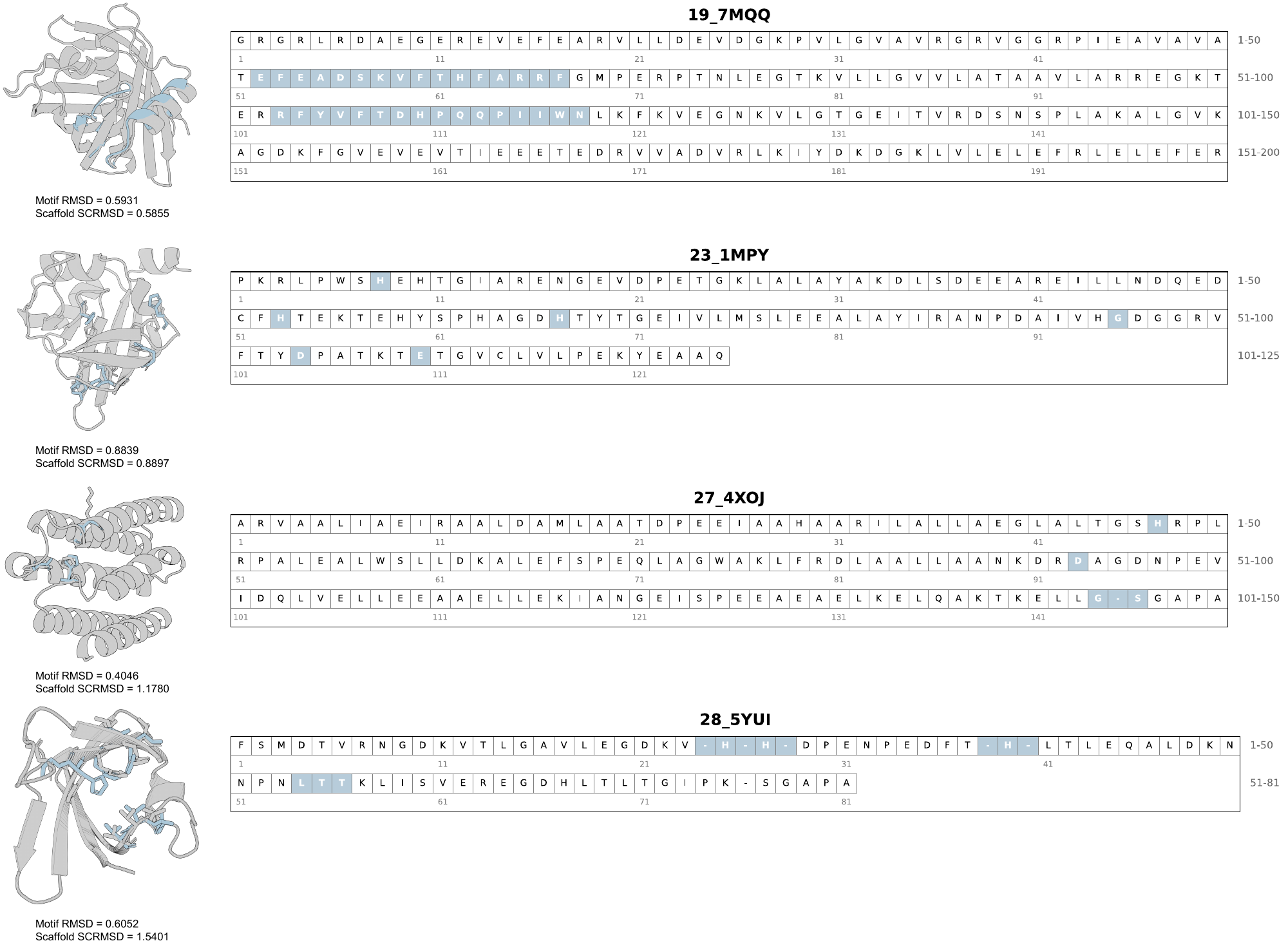}
    \caption{Successfully designed case of MotifBench with ReaPro-1c}
    \label{fig:motifvis}
\end{figure}

\subsection{RFDiffusion Motif Benchmark}
For the backbone design task, we use the same model and the same sampling parameters as the motifbench to better test the generalization ability on different datasets.
For the all-atom design task, we follow the same protocol as motifbench.
We first generate a small set (100 sequences for each motif) to select a checkpoint between 199, 200, and 201 epochs.
Then we use the best checkpoint to obtain the final result.
Note that the baseline model was trained for 383 epochs, which is significantly longer than our model.

\begin{figure}
    \centering
    \includegraphics[width=1\linewidth]{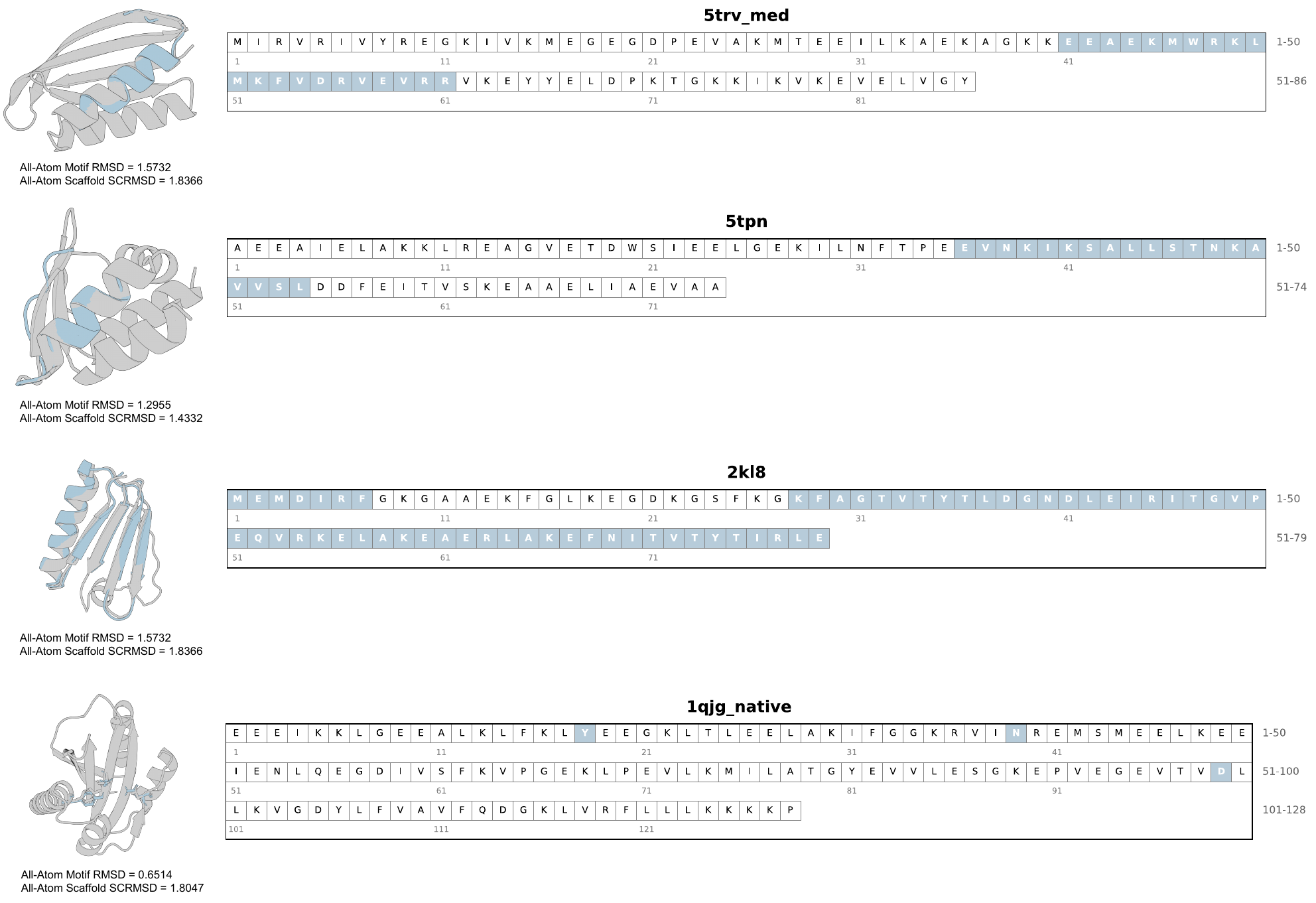}
    \caption{Successfully designed case of RFDiffusion benchmark with ReaPro-1c}
    \label{fig:rfdvis}
\end{figure}

\end{document}